\documentstyle[12pt,psfig]{article}
\input epsf

\begin{document}
\begin{flushright}SUSX-TH/02-007\\
YITP-SB-02-06\end{flushright}
 $\;\;$
          $\;\;$

\newcommand{\nc}{\newcommand}
\nc{\vivi}{very interesting and very important}
\nc{\al}{\alpha}
\nc{\ga}{\gamma}
\nc{\de}{\delta}
\nc{\ep}{\epsilon}
\nc{\ze}{\zeta}
\nc{\et}{\eta}
\newcommand{\th}{\theta}
\nc{\tmn}{\theta_{\mu\nu}}
\nc{\Th}{\Theta}
\nc{\ka}{\kappa}
\nc{\la}{\lambda}
\nc{\rh}{\rho}
\nc{\si}{\sigma}
\nc{\ta}{\tau}
\nc{\up}{\upsilon}
\nc{\ph}{\phi}
\nc{\ch}{\chi}
\nc{\ps}{\psi}
\nc{\om}{\omega}
\nc{\Ga}{\Gamma}
\nc{\De}{\Delta}
\nc{\La}{\Lambda}
\nc{\Si}{\Sigma}
\nc{\Up}{\Upsilon}
\nc{\Ph}{\Phi}
\nc{\Ps}{\Psi}
\nc{\Om}{\Omega}
\nc{\ptl}{\partial}
\nc{\del}{\nabla}
\nc{\be}{\begin{equation}}
\nc{\ee}{\end{equation}}
\nc{\bea}{\begin{eqnarray}}
\nc{\eea}{\end{eqnarray}}
\nc{\ov}{\overline}
\nc{\gsl}{\!\not}
\newcommand{\s}{\mbox{$\sigma$}}
\newcommand{\bi}[1]{\bibitem{#1}}
\newcommand{\fr}[2]{\frac{#1}{#2}}
\newcommand{\gm}{\mbox{$\gamma_{\mu}$}}
\newcommand{\gn}{\mbox{$\gamma_{\nu}$}}
\newcommand{\Le}{\mbox{$\fr{1+\gamma_5}{2}$}}
\newcommand{\R}{\mbox{$\fr{1-\gamma_5}{2}$}}
\newcommand{\GD}{\mbox{$\tilde{G}$}}
\newcommand{\gf}{\mbox{$\gamma_{5}$}}
\newcommand{\Ima}{\mbox{Im}}
\newcommand{\Rea}{\mbox{Re}}
\newcommand{\Tr}{\mbox{Tr}}
\newcommand{\psl}{\slash{\!\!\!p}}
\newcommand{\cp}{\;\;\slash{\!\!\!\!\!\!\rm CP}}
\newcommand{\qq}{\langle \ov{q}q\rangle}
\def\ga{\mathrel{\raise.3ex\hbox{$>$\kern-.75em\lower1ex\hbox{$\sim$}}}}
\def\la{\mathrel{\raise.3ex\hbox{$<$\kern-.75em\lower1ex\hbox{$\sim$}}}}
\nc{\gtwid}{\mathrel{\raise.3ex\hbox{$>$\kern-.75em\lower1ex
\hbox{$\sim$}}}}
\nc{\ltwid}{\mathrel{\raise.3ex\hbox{$<$\kern-.75em\lower1ex
\hbox{$\sim$}}}}

\begin{center}

\large
{\bf Indirect limits on the CPT violating background in the neutrino sector}
\normalsize

\vspace{1cm}

Irina Mocioiu$^1$ 
 and
Maxim Pospelov$^2$

\vspace{1cm}

$^1${\it C.N. Yang Institute for Theoretical Physics\\
State University of New York, Stony Brook, NY 11794-3840}\\ $\;$\\
$^2${\it
         Centre for Theoretical Physics,
                                 CPES,
                                 University of Sussex,
                                 Brighton BN1~9QJ,
                                 United Kingdom}
         
\end{center}


\begin{abstract}
CPT violation in the neutrino sector, suggested as a new way to reconcile
different neutrino anomalies, induces at the radiative level 
observable effects among charged leptons, where high-precision 
tests of the CPT symmetry are available. 
We show that, in the models with 
heavy right-handed Majorana neutrinos, constraints imposed by these 
experiments require CPT violation in neutrino spectrum be suppressed 
to a level undetectable for any conceivable neutrino experiment.
We find that the CPT violation in the neutrino sector 
may evade indirect constraints only at the expense of light 
right-handed neutrinos with small Yukawa couplings to the Standard Model 
sector or by allowing non-locality well below the electroweak scale. 

\end{abstract}

\vspace{4cm}

\hfill\eject

\section{Introduction}
In Quantum Field Theory, any local, hermitian, Poincar\'e invariant action is 
CPT invariant. CPT invariance is thus considered to be a 
fundamental symmetry of particle physics. This does not imply, however,
that the CPT invariance cannot be broken: field theory can be just a low
energy limit of another fundamental theory (e.g., string theory), where 
one or more of the conditions for the CPT theorem are violated 
(e.g., locality). At low energies, field theory still presents  
a perfect description and therefore all effects of CPT violation 
must be described in the form of effective CPT-odd operators. Invariably, 
such operators would break Lorentz invariance, so that the 
breaking of CPT can be ascribed to some background (vector or axial-vector) 
fields that define preferred directions. A classification of such backgrounds,
coupled to the operators of lowest dimensions built from the Standard Model 
fields was given in Ref. \cite{Kost}. Specific 
realizations of the CPT breaking were discussed in the context 
of string theory \cite{KP, Ellis}, in chiral gauge theories on spacetimes
with a particular topological structure \cite{Klink}
and in non-commutative geometries \cite{GAC,MPR}. 
Phenomenologically, CPT violation can be motivated 
by an interesting, although quite speculative idea of equilibrium 
baryogenesis \cite{Kuzmin}.

On the experimental side, there have been diverse efforts 
to detect possible signatures of  
CPT/Lorentz violating backgrounds that resulted in a collection of 
extremely tight bounds on the CPT-odd physics 
\cite{clocks}-\cite{Heckel}. 
It is clear that, up to date, the most precise experiments that check Lorentz
symmetry and CPT where performed with ``friendly'' matter: electrons, 
nuclei, relatively long-lived mesons. Direct constraints on the 
CPT violation in the neutrino sector are by far more modest.

The absence of direct constraints on the CPT violation in neutrino sector 
has sprung a number of interesting speculations that CPT breaking may be 
comparable to  neutrino masses  and splittings between different generations
\cite{MY,L1}. 
The presence of CPT violation creates a  possibility to accommodate 
all neutrino data: atmospheric and solar neutrino anomalies, as well as the 
LSND effect, {\em without} introducing new light degrees 
of freedom such as sterile neutrinos. To account for the LSND 
anomaly, the authors of 
Ref. \cite{L1} split the ``masses'' of electron neutrino and antineutrino 
by $O$(1eV). In a more adequate language, this amounts 
to asymmetrically modifying the 
dispersion relations for neutrino and antineutrino by an eV-size term. 
It is easy to see that, quite generically, these terms can be combined to form 
an effective operator of the form $\bar \nu b_\mu \gamma_\mu (\gamma_5) \nu$. 
Another equally speculative motivation for CPT/Lorentz violation in the 
neutrino sector are the hints on the ``negative'' $m_\nu^2$ 
in the experiments searching for neutrino masses using the end-points of beta 
decays and the seasonal variations of this effect \cite{Lob}.

Considering as an example the models in \cite{MPR}, we would like 
to argue that CPT violation would naturally arise in the sector of 
singlet neutrinos. In the higher dimensional set up of Ref. \cite{MPR},
with non-commutativity between usual 4d and extra-dimensional coordinates, 
the CPT violation is induced for a fermionic Kaluza-Klein zero mode 
if this fermion is allowed to propagate in the bulk. Charged matter 
needs to be chiral, and the projecting out of unwanted states leads to 
CPT even physics \cite{MPR}. On the other hand, singlet 
neutrino is allowed to live in higher dimensions and 
therefore is susceptible to the CPT violating effects. 
 
But even though the neutrino sector is the most likely recipient of CPT 
violation induced by some fundamental physics 
(string theory, non-commutativity, etc.), it is easy to argue from general 
principles that CPT violation cannot reside solely 
in the neutrino sector. The charged leptons and neutrinos are 
connected by electroweak and Yukawa interactions, and therefore CPT 
violation in the neutrino sector will induce CPT violation among charged 
leptons. 
The coupling of the axial vector background to the electron current $\bar e 
\gamma_\mu\gamma_5 e$ is limited to an impressive accuracy \cite{Heckel} 
\be
|\vec b_{{\rm{\small electron}}}|  \la 10^{-28}~ {\rm GeV}
\label{elimit}
\ee
with slightly milder limits coming from the clock comparison experiments
that involve paramagnetic atoms \cite{clocks}.

The purpose of this letter is to study the mechanisms that import 
CPT violation from neutrinos to the charged lepton sector and 
make a full use of the constraint (\ref{elimit}). This allows
to place strong bounds on the CPT-violating neutrino phenomenology and 
narrow down certain theoretical constructions that would evade these
constraints.

\section{CPT-violation with minimal field content}

\begin{figure}
 \centerline{%
   \psfig{file=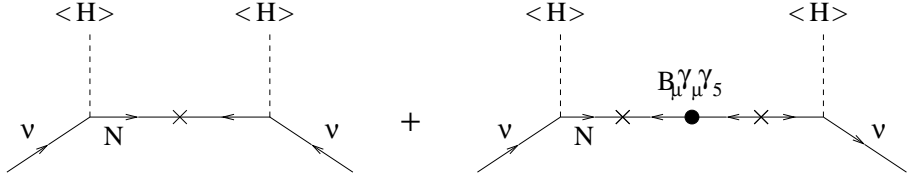,width=12cm,angle=0}%
         }
\vspace{0.1in}
 \caption{Tree level see-saw diagram with the first CPT-odd 
correction. Heavy dot represents the insertion of the CPT-violating 
background and crosses flip the fermion number}
\end{figure}

We start from the models which are the most economical from the point of view 
of the number of light degrees of freedom.  
We assume that at low energies 
($E < M_W$) the neutrino spectrum consists of only three generations of the Standard
Model neutrinos. To have a non-trivial neutrino phenomenology, we 
extend the Standard Model by a set of heavy Majorana neutrinos $N_i$. 
By heavy we mean not necessarily the GUT scale neutrino: for the purposes 
of the present discussion it is sufficient to assume that $M_{i}\ga M_W$. 
Now, we introduce the CPT-violating background in 
the singlet neutrino sector, that would couple to bilinear combinations
of $N$. 

For the three generation case, two types of backgrounds are possible:
\be
{\cal L}_N = \fr{1}{2}N^T C {\bf M} N  + \bar N {\bf A_\mu}
\gamma_\mu  N+ \bar N {\bf B_\mu}
\gamma_\mu \gamma_5 N,
\label{LagN}
\ee 
The matrix ${\bf A_\mu}$ is antisymmetric and ${\bf B_\mu}$ is symmetric in 
generation space. For a single generation case, only the $B_\mu $ term is 
allowed. 
For the clarity of the discussion, we shall work with the first generation and
 extend 
it to the multi-generation case when needed. We further assume that 
the CPT violating terms are generated at some scale $\Lambda_{UV}$. 
It is natural to think that 
this scale is also large, $\Lambda_{UV}\ga M_W$.

Integrating out heavy degrees of freedom, at the tree level, 
we get the standard see-saw 
 mechanism for the left handed neutrinos, i.e. dimension five 
$\Delta L =2$ effective operators, corrected
for the presence of the CPT-violating terms. The relevant tree level 
diagrams are given in Figure 1. 
\be
{\cal L}_\nu=
\nu^TC \fr{(yv)^2}{2M}\nu + \bar\nu \fr{(yv)^2B_\mu\gamma_\mu \gamma_5}
{2(B^2 + M^2)} \nu.
\label{seesaw}
\ee
Here $y$ is the Yukawa coupling and $v=246$ GeV is the electroweak v.e.v.
For future convenience, we also introduce a notation 
\be
b^{{\rm {\small tree}}}_\mu = \fr{(yv)^2B_\mu}
{2(B^2 + M^2)}
\label{btree}
\ee

Now we take into account the loop correction coming from the $H-N$ exchange 
diagram, Figure 2. We choose to work in the explicitly renormalizable gauge 
and 
by $H$ we mean all four scalars that belong to the 
Standard Model Higgs doublet $H$. In the unitary gauge, the exchange by 
charged scalars and a pseudoscalar will be equivalent to taking into account 
longitudinal components of gauge bosons in $W-N$ and $Z-N$ exchange diagrams. 
In the CPT-even channel this loop is not 
important, as it brings a small 
correction to the kinetic term of the left-handed doublet $L$. In the 
CPT-odd channel, the corrections are generated for both the neutral and 
charged components of $L$. 
\be
{\cal L}_L = \bar L \gamma_\mu \gamma_5 L b^{{\rm {\small loop}}}_\mu 
\label{loop}
\ee

\begin{figure}
 \centerline{%
   \psfig{file=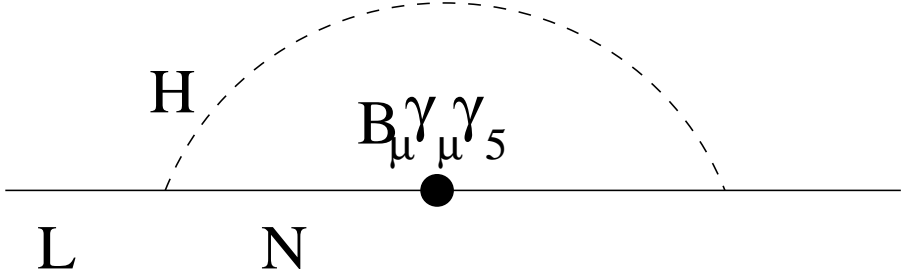,width=12cm,angle=0}%
         }
\vspace{0.1in}
 \caption{$H-N$ loop that generates $\bar L \gamma_\mu \gamma_5 L 
b^{{\rm {\small loop}}}_\mu$ 
interaction }
\end{figure}

Thus, at the loop level, $b^{{\rm {\small loop}}}_{{\rm {\small electron}}}\simeq
b^{{\rm {\small loop}}}_{{\rm {\small neutrino}}}$ and both quantities are 
subject to the constraint (1). Therefore, the only way of having 
the CPT-odd neutrino phenomenology at $O(1$eV$)$-level is to have 
$b^{{\rm {\small loop}}}/b^{{\rm {\small tree}}} < 10^{-19}$!

We now turn to the explicit result for $b^{{\rm {\small loop}}}$:
\begin{eqnarray}
b_\mu^{{\rm {\small loop}}} = \fr{y^2B_\mu}{64\pi^2}\left\{\begin{array}{c}
\ln\fr{\Lambda_{UV}^2}{B^2+M^2}~~{\rm if}~~ \Lambda_{UV}^2 > B^2+M^2\\
c_1 \fr {\Lambda_{UV}^2}{B^2+M^2}~~{\rm if}~~ B^2+M^2>\Lambda_{UV}^2 >M_W^2\\
 c_2\fr {\Lambda_{UV}^4}{M_W^2(B^2+M^2)}~~{\rm if}~~ M_W^2 >\Lambda_{UV}^2
\end{array}\right.
\label{result}
\end{eqnarray}
where the loop integrals were cutoff at $\Lambda_{UV}$, which was kept 
arbitrary and $c_1$ and $c_2$ are order one coefficients. 

It is convenient to rewrite this result in the following form:
\begin{eqnarray}
\fr{b^{{\rm {\small loop}}}}{b^{{\rm {\small tree}}}}
 =  \fr{1}{32\pi^2}\left\{\begin{array}{c}
\fr{B^2+M^2}{v^2}\ln\fr{\Lambda_{UV}^2}{B^2+M^2}~~{\rm if}~~ 
\Lambda_{UV}^2 > B^2+M^2\\
 c_1\fr {\Lambda_{UV}^2}{v^2}~~{\rm if}~~ B^2+M^2>\Lambda_{UV}^2 >M_W^2\\
 c_2\fr {\Lambda_{UV}^4}{M_W^2v^2}~~{\rm if}~~ M_W^2 >\Lambda_{UV}^2
\end{array}\right.
\label{ratio}
\end{eqnarray}
which is valid for every component of $b_\mu$.

Examining the expression (\ref{ratio}), we observe that the first two lines 
are larger than one, so that $b^{{\rm {\small loop}}}\ga 10^{-2}
b^{{\rm {\small tree}}} $ according to our assumptions. This, together with the
experimental bound (1) immediately 
renders the following constraint on the CPT-odd terms in the sector of 
the left-handed neutrinos:
\be
b^{{\rm {\small loop}}}_{{\rm {\small neutrino}}}~~ {\rm and} ~~
b^{{\rm {\small tree}}}_{{\rm {\small neutrino}}}< 10^{-17}{\rm eV}
~~~~~{\rm for}~~ \Lambda_{UV},M>M_W,
\label{bound}
\ee
which is applicable to the spatial components of the CPT-background. 
One may wonder if the bound (\ref{bound}) could be relaxed if 
the four-vector $b$ is chosen to be exactly time-like. If 
$b$ is time-like in the galactic frame, the motion of the solar system 
relative to galactic halo translates into $b_i\sim 10^{-3}b_0$. If this 
effect is also fine-tuned, the motion of the Earth around the Sun breaks it at 
the level of $O(10^{-4})$. In other words, a careful choice of frame for 
$b_\mu$ may relax the bound (\ref{bound}) to the level of $10^{-13}{\rm eV}$,
which is still dramatically lower than the detection possibilities for any 
direct experiments with neutrinos \cite{nucpt}. Would it be possible to have 
eV CPT-odd terms for other flavours, as the CPT violation is by far 
less restricted for muons and tau-leptons? The strength of the bound 
in this case would depend, of course, on the 
mixing angle with the electron neutrinos. However, suppressing 
$\theta_{1i}$ to 
a $O(10^{-8})$ level would also make these neutrinos useless for ``normal''
CPT-even oscillations, making the consistency of all neutrino anomalies even more
problematic. In any event, this possibility will appear again as 
a severe fine-tuning. 
We conclude that the 
minimal scenario - three light neutrinos below the electroweak scale 
and $O($eV$)$-size CPT violation coming from short distances - is 
highly unnatural on account of the strong bounds coming from the tests of
CPT and Lorentz invariance for electrons.

\section{Small $\Lambda_{UV}$ and/or Dirac neutrinos}

While excluding the minimal and the most natural possibilities for 
the CPT-odd neutrino phenomenology, Eqs. (\ref{result}-\ref{bound}) also 
suggest how to suppress the size of effects in the charged lepton sector.

{\em Option 1. Dirac neutrinos: $M\equiv0$, small $y$}

One could consider adding right handed fields with very small Yukawa couplings 
to the Standard Model, so that neutrinos would get small Dirac masses
\be
m_\nu=\frac{y v}{\sqrt{2}}\sim eV
\ee 
In order to obtain the desired CPT-violating 
neutrino phenomenology, $B_\mu$ should also be of the order of $1eV$.
 The effect of CPT violation for electrons is obtained by the same 
radiative mechanism as before, but now it is strongly suppressed by the small
Yukawa couplings.
\be
b_\mu^{{\rm {\small loop}}}=\frac{y^2 B_\mu}{64\pi^2}\ln\frac{\Lambda_{UV}^2}
{M_W^2}=\left (\frac{m_\nu}{v}\right)^2
\frac{B_\mu}{32\pi^2}\ln\frac{\Lambda_{UV}^2}{M_W^2}\sim10^{-23}\frac
{B_\mu}{32\pi^2}\ln\frac{\Lambda_{UV}^2}{M_W^2}
\ee
For a cut-off $\Lambda_{UV}$ of the order of the GUT scale, $\Lambda_{UV}\sim 
2 \times 10^{16 }$GeV, one gets $b_\mu^{{\rm {\small loop}}}\sim 3\times 
10^{-24} B_\mu\sim
 3\times 10^{-33}$GeV. This is about four orders of magnitude below the 
present limit (\ref{elimit}) but might be within the reach of the
 new generation
of clock comparison experiments \cite{Mike}. 
This scenario avoids the constraints from the charged lepton sector, but 
introduces new degrees of freedom far below the electroweak scale and does not
have a natural understanding of the smallness of neutrino masses. 

{\em Option 2. Small values of  $\Lambda_{UV}$ }

Another, more exotic possibility, is that the scale 
at which CPT-odd physics is generated could be much lower than the 
electroweak scale and the loop effects are suppressed as $\Lambda_{UV}^4/v^4$ 
(Eqs. (\ref{result}) and (\ref{ratio})). In order to have loop effects 
under control and the eV-size CPT violation in the neutrino 
sector, one should have $\Lambda_{UV} $ in the 100 KeV range or {\em smaller}. 
Needless to say that this corresponds to a drastically different 
physics for the right-handed neutrino. Without having any convincing 
model for such a possibility, one can imagine an effective CPT-odd 
non-local interaction 
\be
\bar \nu \gamma_\mu\gamma_5B_\mu 
\left(\fr{ \Lambda_{UV}^2}{\partial^2 + \Lambda_{UV}^2}\right)^n \nu
\label{nonlocal}
\ee
with some positive power $n$. Such form of the CPT-odd interaction will 
be seen as a usual axial-vector background at energies smaller than 
$\Lambda_{UV}$, at the same time providing a form factor/cutoff for loop 
integral at the energies larger than $\Lambda_{UV}$. 

This possibility is hard to realize unless one very specific phenomenon 
happens in the right-handed neutrino sector. If the right-handed 
neutrinos are light and very strongly-interacting due to a force of yet 
unknown origin, they might develop a fermion condensate with an open 
fermion number. At low energies a cubic root from the value of 
the corresponding condensate will 
determine $B_\mu$ and at high energies all 
CPT-odd pieces of propagators 
will be power-like suppressed (similarly to 
non-perturbative pieces in high-energy quark and gluon propagators in QCD).

\section{Conclusions}

The main finding of our paper is that the CPT violating effects are very 
efficiently transmitted from neutrinos to the charged lepton sector if the
Majorana right-handed neutrinos and the scale of CPT breaking 
are heavier than the electroweak scale. Using the results of 
\cite{Heckel}, we have shown that the CPT-odd  
energy splitting in the neutrino sector should be suppressed down to at least the 
$10^{-13}$ eV level, which makes it useless for reconciling different neutrino
 anomalies within the minimal model of three light SM neutrinos. 

We see two potential ways of avoiding strong constraints coming from the 
charged lepton sector. Both of them require drastic modifications of 
the theory at the scales {\em lower} than the electroweak scale. 
One possibility is to make right-handed neutrinos light, 
so that corresponding Yukawa couplings
are small, and the radiative mechanism of transmitting CPT violation 
to the charged lepton sector will be suppressed  by an additional 
factor of $(m_\nu/M_W)^2$. But this ``solution'' introduces new degrees 
of freedom far below the electroweak scale and thus defies the purpose 
of having CPT-odd neutrino phenomenology, which was invented 
as an alternative to sterile neutrino models.  

Another possibility is even more exotic. The CPT violating terms 
should have a ``form factor'', i.e. the theory becomes effectively 
non-local at the scale of 1 GeV or less. Finally, both possibilities 
can be merged together if the right-handed neutrinos are light and 
for some reasons develop a condensate with a non-zero fermion number. 
It will be a serious model-building 
challenge to find a model for right-handed neutrinos that would 
furnish this property. 

As a concluding remark, we would like to mention that if one of this 
possibilities is realized and CPT violation in neutrino sector escapes 
indirect constraints (however unlikely it looks at the moment), 
the CPT-odd neutrino phenomenology can be highly anisotropic and 
oscillation patterns can be different from what was discussed in 
\cite{MY,L1,nucpt}. Previous discussions
of CPT odd effects effectively concentrated on the time-like vectors 
$b_\mu \sim (b_0,0,0,0)$. At the  moment
there are no solid reasons to argue that the space components of $b$-vector 
should be zero. This brings a question of directional dependence  
and sidereal variations in CPT-odd neutrino phenomenology that come 
atop of already known anisotropies in CPT-even neutrino physics 
such as azimuthal 
dependence of atmospheric neutrino fluxes, day/night effects, etc.

\end{document}